\documentclass[a4paper]{article}
 \usepackage{booktabs}
 \usepackage[thinc]{esdiff}
 \usepackage{multirow,xcolor}
\usepackage{INTERSPEECH2020}
\newcommand{\be}{\begin{eqnarray}}
\newcommand{\ee}{\end{eqnarray}}

\title{Deep Learning Based Dereverberation of Temporal Envelopes \\ for Robust Speech Recognition}
\name{Anurenjan Purushothaman, Anirudh Sreeram, Rohit Kumar, Sriram Ganapathy\thanks{This work was funded by the project grants from Samsung Research India, Bangalore.}}
\address{
  Learning and Extraction of Acoustic Patterns (LEAP) lab,\\
  Electrical Engineering, Indian Institute of Science, Bangalore.\email{\{anurenjanr, sanirudh, rohitk, sriramg\}@iisc.ac.in}}

\begin{document}

\maketitle
\begin{abstract}
   Automatic speech recognition in reverberant  conditions is a challenging task as the long-term envelopes of the reverberant speech are temporally smeared. 
   In this paper, we propose a neural model for enhancement of sub-band  temporal envelopes for dereverberation of speech. The  temporal envelopes are derived using the autoregressive modeling framework of frequency domain linear prediction (FDLP).  The neural enhancement model proposed in this paper performs an envelop gain based enhancement of temporal envelopes and it consists of a series of convolutional and recurrent neural network layers. The enhanced sub-band envelopes are used to generate features for automatic speech recognition (ASR). 
   %Even when the enhancement model is trained only on simulated reverberation, we show that this model can generalize to real reverberant data. 
   The ASR experiments are performed on the REVERB challenge dataset as well as the CHiME-3 dataset. In these experiments, the proposed neural enhancement approach provides significant improvements over a baseline ASR system with beamformed audio (average relative improvements of $21$\% on the development set and about $11$\% on the evaluation set  in word error rates for REVERB challenge dataset).
   \end{abstract}

\section{Introduction}

Automatic speech recognition (ASR) systems find widespread use in applications like human-machine interface, virtual assistants, smart speakers etc, where the input speech is often reverberant and noisy. The ASR performance has improved dramatically over the last decade with the help of deep learning models \cite{yu2016automatic}. However, the degradation of the systems in presence of noise and reverberation continues to be a challenging problem due to the low signal to noise ratio \cite{hain2012transcribing}.  For \textit{e.g.} Peddinti \textit{et al.,} \cite{dan} reports a $75\%$ rel. increase in word error rate (WER) when signals from a far-field array microphone are used in place of those from headset microphones in the ASR systems, both during training and testing. This degradation could be primarily attributed to reverberation artifacts which smear the time domain envelopes of the speech signal \cite{yoshioka2012making, reverb}. 
%The availability of multi-channel signals can be leveraged for alleviating these issues as most of the real life far-field speech recordings are captured by a microphone array.

\begin{figure*}[t!]
  \centering
  \includegraphics[scale=0.33]{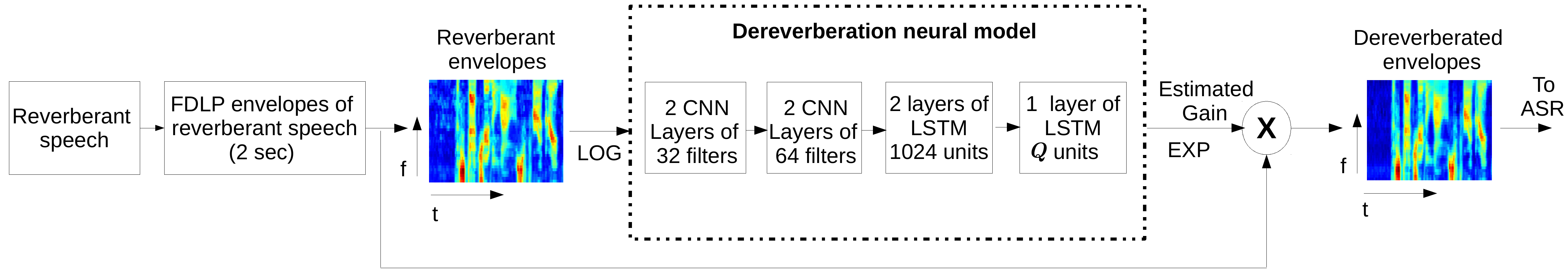}
      \caption{Block schematic of envelope dereverberation model}
  \label{fig1}
\end{figure*}

%Previously, many works have focused on far-field speech recognition using multiple microphones \cite{far1,far2,far3,far4}. 
The  traditional approach to multi-channel far-field ASR combines all the available channels by beamforming \cite{anguera2007acoustic}. Recently,  unsupervised DNN-mask estimator based beamforming is also proposed for generalized eigen value (GEV) based beamforming \cite{rohit}. Along with the beamforming, the weighted prediction error (WPE)~\cite{wpe} based dereverberation is used in state-of-art ASR systems in reverberant environments. In addition, multi-condition training, where reverberation is simulated in training data is commonly employed to reduce the mis-match between training and testing \cite{seltzer2013investigation}.  However, even with these methods, the temporal smearing of sub-band envelopes, caused by the combination of the direct path and the reflected paths in reverberation, continue to degrade the ASR performance  \cite{peddinti2017low}.

In this paper, we propose an approach for sub-band envelope enhancement which  attempts to learn the mapping of the reverberated envelopes to the close-talking ones. The sub-band envelopes are derived using the autoregressive modeling framework of frequency domain linear prediction \cite{thomas2008recognition,ganapathy2018far}.  A deep neural model based on convolutional and recurrent layers is trained to enhance the reverberated sub-band FDLP envelopes. Following the DNN model training, which predicts an envelope gain, the output of the model is multiplied with the sub-band envelopes of the reverberant speech to suppress the effects of reverberation. The enhanced sub-band envelopes are used for feature extraction of ASR. In various ASR experiments on the REVERB challenge dataset \cite{reverb} as well as the CHiME-3 dataset \cite{chime3}, we show that the proposed approach improves over the state-of-art ASR systems based on log-mel features with GEV beamforming and WPE enhancement. 

%model for level enhancement of frequency domain linear prediction (FDLP) based features. FDLP is the frequency domain dual of Time Domain Linear Prediction (TDLP). Just as TDLP estimates the spectral envelope of a signal, FDLP estimates the temporal envelope of the signal, i.e. square of its Hilbert envelope \cite{analytic}. The frame work relies on frequency domain linear prediction which states that, linear prediction applied on frequency domain estimates the envelopes of the signal \cite{athineos2003frequency,sriphd}.

%A CNN-LSTM network is used to enhance the FDLP features by training with simulated data from available data sets with mean square error between the simulated reverberant and noisy speech and the corresponding clean target, as the training criteria.  With several ASR experiments conducted on CHiME-3 \cite{chime3} and REVERB challenge dataset \cite{rev1,rev2}, we show that the proposed enhancement approach  improves significantly  over a baseline system using conventional WPE applied DNN-mask based beamformed audio with mel filter bank energy features.

%The rest of the paper is organized as follows. The related prior work is  discussed in Section~\ref{sec:prior_work}. The details about the proposed dereverberation approach are provided in Section~\ref{sec:fdlp_enhan}. The ASR experiments and results are reported in Section~\ref{sec:expt}, which is followed by a  summary in Section~\ref{sec:summary}. 

\section{Related Prior Work}\label{sec:prior_work}
%Speech enhancement based on neural networks has made noticeable progress in the recent years. 
The early works by Xu et. al. \cite{xu2014regression} targeted the enhancement of signals corrupted by additive noise where a supervised neural network method was proposed to enhance speech by means of finding a mapping function between noisy and clean speech signals. In a similar manner, speech separation (the problem of separating the target speaker speech from the background interference) has seen considerable progress using neural methods with ideal ratio mask based mapping \cite{wang2018supervised}.

For reverberant speech, Zhao et al., proposed a LSTM model to predict late reflections in the spectrogram domain \cite{zhao2018late}. A spectral mapping approach using the log-magnitude inputs was attempted by Han et. al \cite{han2014learning}. A mask based approach to dereverberation on the complex short-term Fourier transform domain was explored by Williamson et. al \cite{williamson2017time}. A recurrent neural network model to predict the spectral magnitudes for dereverberation of speech was also proposed by Santos et. al \cite{santos2018speech}. 
Speech enhancement for speech recognition based on neural networks has been explored in \cite{wollmer2013feature,chen2015speech,weninger2015speech}. In \cite{maas2013recurrent} a recurrent neural network is used to  map  noise-corrupted  input  features  to  their  corresponding  clean  versions.

%Weighted prediction error (WPE) based reverberation suppression and beamforming techniques like delay and sum beamforming \cite{anguera2007acoustic}, generalized eigen value beamformer \cite{gev} are aiming to reduce the effect of reverberation and at the same time trying to increase the signal to noise ratio. These methods directly try to enhance the speech signal.

%In \cite{krueger2010model}, a model based enhancement of the logarithmic Mel power spectrum for reverberant speech is dealt with. Deng et al. used speech feature enhancement for dynamic compensation of hidden Markov model (HMM) variances, showing better recognition accuracy in \cite{deng2005dynamic}.

%In \cite{athineos2003frequency}, Athenios et al. came up with features based on temporal envelopes, extracted using frequency domain linear prediction (FDLP), which are robust in reverberant conditions \cite{thomas}. Separately, a multi-band feature extraction using autoregressive modeling was proposed for deriving noise robust features from single channel speech \cite{marsri,ganapathy2018far}.

%In this paper, we propose a CNN-LSTM neural network for cleaning the robust sub-band temporal envelopes proposed in \cite{sriphd}.

\section{Proposed Approach}\label{sec:fdlp_enhan}
\subsection{Signal model}\label{sec:signal_model}
When speech is recorded in far-field reverberant environment, the data collected in the microphone can be expressed as 
\be
\label{eq:reverb_sig}
r(t) = x(t)*h(t),
\ee
where $x(t)$, $h(t)$ and $r(t)$  denote the clean speech signal, the room impulse response and the reverberant speech respectively. The room response function $h(t) = h_e(t) + h_l(t)$, where $h_e(t)$ and $h_l(t)$ represent the early and late reflection components. 

Let $x_q(t)$, $h_q(t)$ and $r_q(t)$ denote the sub-band clean speech, room-response and the reverberant speech respectively and $q=1,~...~,Q$ denotes the sub-band index. Assuming an ideal band-pass filtering we can write (using Eq.~\ref{eq:reverb_sig}),
\be
r_q(t) = x_q(t)*h_q(t).
\ee
Now, the analytic signal $r_{aq}(t) = r_q(t) + \mathcal {H} [r_q(t)]$ can be shown to be \cite{thomas2008recognition,ganapathy},
\be
\label{eq:reverb_analytic_conv}
r_{aq}(t) = \frac{1}{2}[x_{aq}(t)*h_{aq}(t)],
\ee
For band-pass filters with small band-width, applying magnitude on both sides, we get the following approximation between the sub-band envelope (defined as the magnitude of the analytic signal) components of the reverberant signal and those of the clean speech signal.
\be
\label{eq:envelope_conv_model}
m_{rq}(t) \simeq \frac{1}{2} m_{xq}(t)*m_{hq}(t),
\ee

where $m_{rq}(t)$, $m_{xq}(t)$, $m_{hq}(t)$ denote the sub-band envelopes of reverberant speech, clean speech and room response respectively. With this model of reverberation in the envelope domain, we can further split the envelope into  early and late reflection coefficients. 
\be
\label{eq:envelope_conv_model_early}
m_{rq}(t) = m_{rqe}(t) + m_{rql}(t),
\ee
In this work, the envelopes are also estimated using the autoregressive modeling framework of frequency domain linear prediction (FDLP). Specifically, the discrete cosine transform (DCT) of sub-band signals $r_q(t)$ is computed and a linear prediction (LP) is applied on the DCT components. The LP envelope estimated using the prediction on the DCT components provides an all-pole model of the sub-band envelopes $m_{rq}(t)$ \cite{ganapathy}.

%Frequency domain linear prediction (FDLP) is the dual of time domain linear prediction. By applying linear prediction in the frequency domain, we can extract the temporal envelopes of a signal. If we are applying linear prediction in a specific sub-band, we get the temporal envelope corresponding to that sub-band. Here, we are using a CNN-LSTM based neural network to enhance the FDLP envelopes, thus obtained. The CNN-LSTM network is trained with paired reverberant and noisy speech and the corresponding clean speech with mean square error loss criteria.
%\subsection{Frequency Domain Linear Prediction}
%FDLP is the frequency domain dual of Time Domain Linear Prediction (TDLP). Just as TDLP estimates the spectral envelope of a signal, FDLP estimates the temporal envelope of the signal, i.e. square of its Hilbert envelope \cite{analytic}. Temporal envelope is given by the inverse Fourier transform of the auto-correlation function of DCT,
%\begin{equation}
%    e(t) = F^{-1}\left %\{Autocorr(y[k])\right \}
%\end{equation}
%where $y[k]$ is the DCT of a signal $x[n]$ having $N$- points. The auto-correlation of the DCT signal is defined as,
%\begin{equation}
%    r_y[\tau]=\frac{1}{N}\sum_{k=\lver%t\tau\rvert}^{N-1}y[k]y[k-\lvert\tau\rvert]
%\end{equation}
% We use the auto-correlation of the DCT coefficients to predict the temporal envelope of the signal. One of the inherent property of linear prediction is that, it tries to approximate the peaks very well. The FDLP model  tries to preserve the peaks in temporal domain.
\begin{figure*}[t!]
  \centering
  \includegraphics[scale=0.55]{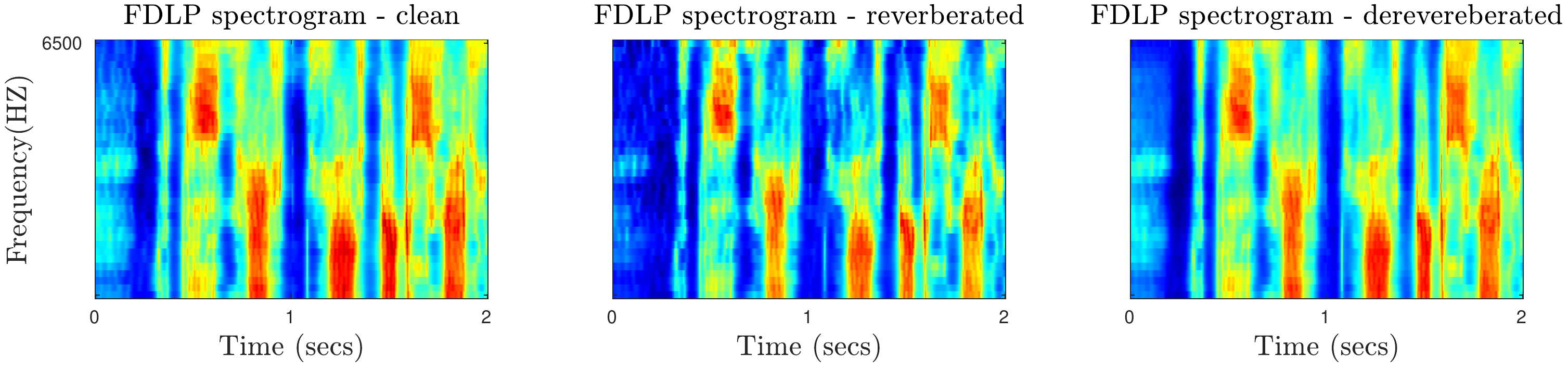}
      \caption{Comparison of spectrograms, FDLP spectrogram for clean (near-room), reverberant speech (far-room) and far-room after the proposed dereverberation, recordings from the REVERB Challenge dataset.}
      
  \label{fig:SSAR_Spectrogram}
\end{figure*}
% \begin{figure*}[t!]
%   \centering
%   \includegraphics[scale=0.56]{icassp_plot_env.pdf}
%       \caption{Comparison of temporal envelopes, FDLP envelopes for clean (near-room), reverberant speech (far-room) and far-room after the proposed dereverberation, recordings from the REVERB Challenge dataset.}
%       \vspace{-0.5cm}
%   \label{fig:env_enhancement}
% \end{figure*}

\begin{figure}[t!]
  \centering
   \includegraphics[scale=0.45]{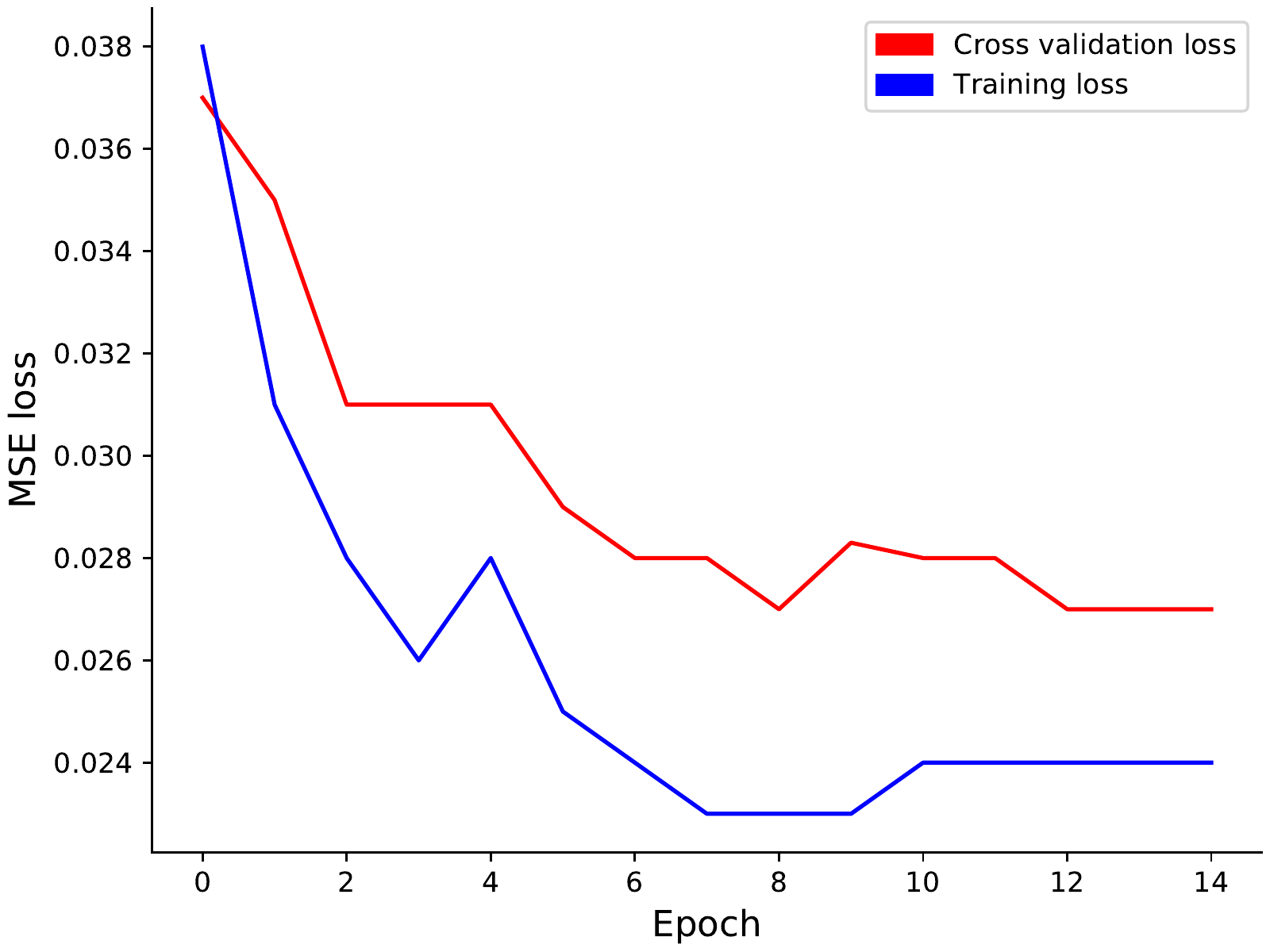}
  \caption{MSE loss for REVERB challenge dataset}
    
  \label{fig:MSE_loss_rev}
\end{figure}
%\begin{figure}[t!]
%  \centering
%  \includegraphics[scale=0.47]{training_cv.png}
%   \includegraphics[scale=0.47]{chime3train_cv.pdf}
 %     \caption{MSE loss for CHiME-3 dataset}
  %\label{fig:MSE_loss_rev}
%\end{figure}

\subsection{Envelope dereverberation model}
As seen in Eq.~(\ref{eq:envelope_conv_model_early}), the FDLP envelope of reverberant speech can be expressed as sum of the direct component (early reflection) and those with the late reflection. In the envelope dereverberation model, our aim is to input the envelope of the reverberant sub-band temporal envelope $m_{rq}(t)$ to predict the late reflection components $m_{rql}(t)$. Once this prediction is achieved, the late reflection component can be subtracted from the sub-band envelope to suppress the artifacts of reverberation. A similar analogy to this envelope subtraction approach is the spectral subtraction model where the noise and clean power spectral density (PSD) gets added in noisy speech PSD. If Gaussian assumptions are made for PSD components \cite{martin2005speech}, the Wiener filtering approach to noisy speech enhancement provides the minimum mean squared error, where the noisy PSD is multipled by the gain of the filter. In a similar manner,  we pose the dereverberation problem as an envelope gain estimation problem. The sub-band envelope gain in this case is the ratio of the sub-band envelope for the direct components to the sub-band envelope of the reverberant sub-band signal. This sub-band envelope gain estimation is achieved using a deep neural network model in the proposed work. Following the model training, the dereverberation is achieved by multiplying the estimated sub-band envelope gain with the sub-band envelope of reverberant speech.

\subsection{Implementation of the envelope dereverberation}\label{sec:neural_network}
The block schematic of the envelope dereverberation model is shown in Figure~\ref{fig1}. The input to the dereverberation model is the FDLP sub-band envelope of the reverberant speech. The model is trained to learn the sub-band envelope gain which is the ratio of the clean envelopes (direct component) with the reverberant envelopes. We use the FDLP envelope of the close talking microphone as an estimate of the direct component. As the envelopes and the gain parameters are positive in nature, the model implementation in the neural architecture uses a logarithmic transform at the input and the estimated gain is followed by an exponential operation. This implementation in the log envelope domain makes the model behave like a residual network based dereverberation architecture. It is also noteworthy that the entire model developed in Section~\ref{sec:signal_model} is applicable only on long analysis windows (which are typically greater than the T60 of the room response function). Hence, unlike previous models for dereverberation, the proposed approach operates on long temporal envelopes of the order of $2$ sec. duration. In the neural model, we also predict the envelope gain of all sub-bands jointly to exploit the sub-band correlations that exist in speech.  

%After applying pre-processing steps like, WPE and unsupervised DNN-mask estimator based GEV beamforming, we get a single channel signal from the available multi-channel far-field recordings. 
From the reverberant speech and the corresponding clean speech, the FDLP sub-band envelopes corresponding to $2$sec. non-overlapping segments are extracted. The choice of non-overlapping $2$sec. is due to computational considerations and the T60 values encountered in practice. If the input sampling rate is $16$ kHz, a $2$sec. segment will correspond to 32,000 samples. FDLP envelopes are extracted at a down sampled rate of $400$ Hz. Thus every $2$sec. segment of audio corresponds to $800$ samples of FDLP envelope for each sub-band. We use a $36$ band mel decomposition. This makes the representation at the input of the enhancement model of size $800 \times 36$. 
The target signal for the enhancement model in Figure~\ref{fig1} is the ratio of the close talking (clean) FDLP envelopes with those of reverberant envelopes.  

The architecture of the neural model is based on convolutional long short term memory (CLSTM) networks (Figure~\ref{fig1}). The input $2$-D data of sub-band envelopes are fed to a set of convolutional layers where the first two layers have $32$ filters each with kernels of size of $41 \times 5$. The next two CNN layers have $64$ filters with $21 \times 3$ kernel size. All the CNN layer outputs with ReLU activations are zero padded to preserve the input size and no pooling operation is performed.  The output of the CNN layers are reshaped to perform time domain recurrence using $3$ layers of LSTM cells. The first two LSTM layers have $1024$ cells while the last layer has $36$ cells corresponding to the size of the target signal (envelope gain). The training criteria is based on the mean square error between the target and predicted output. The model is trained with stochastic gradient descent using Adam optimizer. The dereverberated envelopes are integrated into $25$ms windows with a shift of $10$ ms and these are log transformed and used as features for ASR \cite{ganapathy2012signal}.

An analysis of dereverberation training loss variation as a function of the epoch is shown in Figure~\ref{fig:MSE_loss_rev}. The training loss and validation loss show consistent reduction during the training process. We run the dereverberation model for about $10$ epochs.

An illustration of the envelope enhancement is shown in Figure~\ref{fig:SSAR_Spectrogram}. Here, we plot the FDLP spectrogram (integrated envelopes) for clean signal, reverberated signal and enhanced signal (using the dereveberation model). As seen in Figure~\ref{fig:SSAR_Spectrogram}, the dereverberation model improves the spectogram visibly and makes it closer to the clean FDLP spectrogram. 

%A more careful visualization of the dereverberation can be achieved using the plot of the sub-band envelope of one single sub-band ($10$ th mel-band) as shown in Figure~\ref{fig:env_enhancement}. The sub-band envelopes of reverberant signal deviate from their clean signal counterparts (as explained in Sec.~\ref{sec:signal_model}). Using the dereverberation model proposed in this paper, we find that the FDLP envelopes are more closely matched with the clean signal envelopes.    

\section{Experiments and results}\label{sec:expt}
The experiments are performed on REVERB challenge and CHiME-3 datasets. For the baseline model, we use WPE enhancement along with unsupervised GEV beamforming. This signal is processed with filter-bank energy features (denoted as BF-FBANK). The FBANK features are $36$ band log-mel spectrogram with frequency range from $200$ Hz to $6500$ Hz. This is the same frequency decomposition used in the FDLP and FDLP-dereverberation experiments. The acoustic model corresponds to 2-D CLSTM network described in \cite{anu1}, consisting of 4 layers of CNN, a layer of LSTM with 1024 units performing recurrence over frequency and 3 fully connected layers with batch normalization.

\subsection{ASR framework}
 We used Kaldi toolkit \cite{kaldi} for deriving the senone alignments used in the PyTorch deep learning framework for acoustic modeling. A hidden Markov model - Gaussian mixture model (HMM-GMM) system is trained with MFCC (Mel Frequency Cepstral Coefficients) features \cite{mfcc} to generate the alignments for training the CLSTM model. A tri-gram language model \cite{trigram} is used in the ASR decoding and the best language model weight  obtained from development set is used for the  evaluation set.

\subsection{REVERB Challenge ASR}
The REVERB challenge dataset \cite{rev3} for ASR consists of $8$ channel recordings with real and simulated reverberation conditions. The  simulated data is comprised of reverberant utterances generated (from the WSJCAM0 corpus \cite{rev1}) obtained by artificially convolving clean WSJCAM0 recordings with the measured room impulse responses (RIRs) and adding noise at an SNR of $20$ dB. The simulated data has six different reverberation conditions. The real data, which is comprised of utterances from the MC-WSJ-AV corpus \cite{rev2}, consists of utterances spoken by human speakers in a noisy reverberant room. The training set consists of $7861$ utterances from the clean WSJCAM0 training data by convolving with $24$ measured RIRs.
\begin{table}[t!]
\caption{Word Error Rate (\%) in REVERB dataset for different features and proposed dereverberation method.}
\resizebox{\columnwidth}{!}{
%\centering
\begin{tabular}{@{}l|ccc|ccc@{}}
\toprule
\multicolumn{1}{c|}{\multirow{2}{*}{\textbf{\begin{tabular}[c]{@{}c@{}}Model\\ Features\end{tabular}}}} & \multicolumn{3}{c|}{\textbf{Dev}}                                              & \multicolumn{3}{c}{\textbf{Eval}}                                             \\ \cmidrule(l){2-7} 
\multicolumn{1}{c|}{}                                                                                   & \multicolumn{1}{l}{\textbf{Real}} & \multicolumn{1}{l}{\textbf{Simu}} & \multicolumn{1}{l|}{\textbf{Avg}} & \multicolumn{1}{l}{\textbf{Real}} & \multicolumn{1}{l}{\textbf{Simu}} & \multicolumn{1}{l}{\textbf{Avg}} \\ \midrule
BF-FBANK                                                                                          & 19.1                     & 6.1                      & 12.6                     & 14.7                     & \textbf{6.5}                      & 10.6                    \\
BF-FDLP                                                                                           & 17.8                     & 6.8                      & 12.3                     & 14.0                     & 7.0                      & 10.5                    \\
BF-FBANK + derevb.                                                                                            & 17.3                     & 5.5                      & 11.4                     & 13.1                     & 6.9                      & 10.0                    \\
BF-FDLP + derevb.                                                                                           & \textbf{14.4 }                    & \textbf{5.3 }                     & \textbf{9.9}                      & \textbf{12.0}                     & 6.8                      & \textbf{9.4}                     \\ \bottomrule
\end{tabular}}
\label{table:1}
\end{table}
\subsubsection{Discussion}
Table \ref{table:1} shows the WER results for experiments on REVERB challenge dataset. The WPE applied unsupervised GEV beamformed signal is used for  the FDLP baseline (denoted as BF-FDLP). The BF-FDLP baseline by itself is better than the BF-FBANK baseline (average relative improvements of $2$\% on the development set and about $1$\% on the evaluation set). For a fair comparision of the proposed approach, we have applied a similar dereverbaration method on BF-FBANK baseline. Here, we have trained the neural model with log-mel features corresponding to $2$ sec. duration with all the $36$ mel-bands jointly. This approach is denoted as BF-FBANK + dereverberation. Average relative improvements of $10$\% on the development set and about $6$\% on the evaluation set is achieved compared to the BF-FBANK baseline.

Finally applying the proposed neural model based dereverberation on BF-FDLP baseline (denoted as BF-FDLP + dereverberation) yields average relative improvements of $21$\% on the development set and about $11$\% on the evaluation set, compared to the BF-FBANK baseline. The improvement in real condition is much more than that of simulated data. Average relative improvements of $25$\% on the real development set and about $18$\% on the real evaluation set, compared to the BF-FBANK baseline is achieved by the proposed method. This suggests that, even though the neural model is trained only with simulated reverberations, it generalizes well on unseen real data. 

\subsection{CHiME-3 ASR}
The CHiME-3 dataset \cite{chime3} for the ASR has multiple microphone tablet device recording in four different environments, namely, public transport (BUS), cafe (CAF), street junction (STR) and pedestrian area (PED). For each of the above  environments real and simulated data are present. The real data consists of $6$ channel recordings from WSJ0 corpus sampled at $16$ kHz spoken in the four varied environments. The simulated data was constructed by mixing clean utterances with the environment noise. The training dataset consists of $1600$ (real) noisy recordings and $7138$ (simulated) noisy recordings from $83$ speakers.

\begin{table}[t!]
\caption{Word Error Rate (\%) in CHiME-3 dataset for different features and proposed dereverberation method.}
\resizebox{\columnwidth}{!}{
%\centering
\begin{tabular}{@{}l|ccc|ccc@{}}
\toprule
\multicolumn{1}{c|}{\multirow{2}{*}{\textbf{\begin{tabular}[c]{@{}c@{}}Model\\ Features\end{tabular}}}} & \multicolumn{3}{c|}{\textbf{Dev}}                                              & \multicolumn{3}{c}{\textbf{Eval}}                                             \\ \cmidrule(l){2-7} 
\multicolumn{1}{c|}{}                                                                                   & \multicolumn{1}{l}{\textbf{Real}} & \multicolumn{1}{l}{\textbf{Simu}} & \multicolumn{1}{l|}{\textbf{Avg}} & \multicolumn{1}{l}{\textbf{Real}} & \multicolumn{1}{l}{\textbf{Simu}} & \multicolumn{1}{l}{\textbf{Avg}} \\ \midrule
BF-FBANK & 7.8 & 8.0 & 8.0 & 14.0 & 9.7 & 11.8 \\
BF-FDLP  & 7.0 & 8.1 & 7.5 & {12.0} & 10.0 & 11.0 \\
BF-FBANK + derevb.   & 7.2     & 8.3     & 7.7   & 12.9     & 9.8    & 11.4     \\
BF-FDLP + derevb.  & 7.2    & \textbf{7.9}    & 7.5    & 13     & \textbf{9.6}    & 11.3     \\ 
$~~~~~$ + reg.  & \textbf{6.9 }   & 8.0    & \textbf{7.4}    & \textbf{11.8}     & 9.8    & \textbf{10.8}     \\ \bottomrule
\end{tabular}}

\label{table:2}
\end{table}
\vspace{0.25cm}
\begin{table}[t!]
\caption{Word Error Rate (\%) in CHiME-3 dataset for different features and proposed dereverberation method with RNN-LM}
\resizebox{\columnwidth}{!}{
%\centering
\begin{tabular}{@{}l|ccc|ccc@{}}
\toprule
\multicolumn{1}{c|}{\multirow{2}{*}{\textbf{\begin{tabular}[c]{@{}c@{}}Model\\ Features\end{tabular}}}} & \multicolumn{3}{c|}{\textbf{Dev}}                                              & \multicolumn{3}{c}{\textbf{Eval}}                                             \\ \cmidrule(l){2-7} 
\multicolumn{1}{c|}{}                                                                                   & \multicolumn{1}{l}{\textbf{Real}} & \multicolumn{1}{l}{\textbf{Simu}} & \multicolumn{1}{l|}{\textbf{Avg}} & \multicolumn{1}{l}{\textbf{Real}} & \multicolumn{1}{l}{\textbf{Simu}} & \multicolumn{1}{l}{\textbf{Avg}} \\ \midrule
BF-FBANK & 5.8 & 6.2 & 6.0 & 10.8 & 7.2 & 9.0 \\
BF-FDLP  & 5.1 & 6.1 & 5.6 & \textbf{9.2} & 7.5 & \textbf{8.4} \\
BF-FBANK + derevb.   & 5.0     & 6.3     & 5.7   & 10.0     & 7.6    & 8.8     \\
BF-FDLP + derevb.  & 5.0    & 6.2    & 5.6    & 9.9     & \textbf{7.2}    & 8.6     \\ 
$~~~~~$ + reg.  & \textbf{5.0}   & \textbf{6.0}    & \textbf{5.5}    & 9.5     & 7.6    & 8.6     \\ \bottomrule
\end{tabular}}
\label{table:3}
\end{table}

\subsubsection{Discussion}
The WER results for experiments on CHiME-3 dataset are shown in Table \ref{table:2}. The FDLP baseline, denoted as BF-FDLP is better than the FBANK baseline (BF-FBANK). We observe  average relative improvements of $8$\% on the development set and about $12$\% on the evaluation set when comparing BF-FDLP and BF-FBANK baseline systems. It can also be seen from Table \ref{table:2} that the proposed dereverberation method improves the FBANK-baseline system. Table \ref{table:3} shows the results with recurrent neural network based language model (RNN-LM). 

In the CHiME-3 dataset, we observed that the significant cause of degradation in the signal quality came from the additive noise sources. Hence, the application of the dereverberation model degraded the performance on the BF-FDLP system (which showing improvements in the BF-FBANK system). On further investigation, we found that the dereverberation model also resulted in smoothing of the spectral variations in the FDLP spectrogram. In order to circumvent this issue, we regularized the MSE loss with a term that encouraged the spectral channels to be uncorrelated. The regularization parameter was kept at $0.05$. Using this regularized MSE loss, we improved the BF-FDLP-Dereverberation system results over the dereverberation approach with MSE loss alone. We performed a statistical significance test \cite{stat_sign} comparing the decoded outputs of the proposed systems with the decoded outputs of the baseline system. In this analysis, the probability of improvement of the proposed dereverberation system with regularization (BF-FDLP + derevb. + reg.)  over the FDLP baseline (BF-FDLP) is above $95$\% on all test conditions in real and simulated. These experiments suggest that, even for the audio data  without significant late reflection components (like CHiME-3 dataset), the proposed approach improves significantly over the baseline method (average relative improvements of $8.5$ \% over the baseline BF-FBANK system in the eval. condition). 

% Although, the neural model based dereverberation approach on log-mel features (denoted as BF-FBANK + dereverberation) is showing some improvement compared with baseline,  BF-FBANK, the FDLP envelope based dereverberation (deonted as BF-FDLP + dereverberation) takes a hit when compared to the BF-FDLP baseline. This may be due to the fact that the CHiME-3 data is predominantly noise than reverberation and the use of MSE loss as the loss function also introduces spectral smoothing, which degrades the performance of the ASR. In order to alleviate this problem, we have introduced a regularizing term which is proportional to the derivative of the predicted gain along frequency axis. Let the MSE loss, $L_{m}$ be given by,
% \begin{equation}
%     L_{m} = \sum_{q,t}[m_{rql}(t) - \hat{m}_{rql}(t)]^2
% \end{equation}
% where $m_{rql}(t)$, $\hat{m}_{rql}(t)$ are the target gain and predicted gain for a particular sub-band $q$ and time instant $t$ respectively. The new loss after regularization, $L_{m}^{r}$ is given by,
% \begin{equation}
%     L_{m}^{r} = L_{m} + \lambda \times \diffp{m_{rql}(t)}{q}
% \end{equation}
% where $\lambda$ is a control parameter. The WER result with the new regularization for a control parameter value of $0.1$ is shown in the table \ref{table:2} as, BF-FDLP + dereverberation + regu. As seen, the results are comparable to the BF-FDLP baseline. Thus the proposed dereverberation approach is not hurting in cases where noise is dominant compared to reverberation.

\section{Summary}\label{sec:summary}
In this paper, we propose a new neural model for dereverberation of temporal envelopes. Using the proposed neural dereverberation approach, we perform speech recognition experiments on the  REVERB challenge dataset as well as on the CHiME-3 dataset. These experiments indicate that the proposed neural dereverberation approach generalizes well on unseen reverberant data. The analysis of results also highlight the incremental benefits achieved for application of the proposed approach in other features like, log-mel filter bank features. 
\newpage

\bibliographystyle{IEEEtran}

\bibliography{mybib,refs}

% \begin{thebibliography}{9}
% \bibitem[1]{Davis80-COP}
%   S.\ B.\ Davis and P.\ Mermelstein,
%   ``Comparison of parametric representation for monosyllabic word recognition in continuously spoken sentences,''
%   \textit{IEEE Transactions on Acoustics, Speech and Signal Processing}, vol.~28, no.~4, pp.~357--366, 1980.
% \bibitem[2]{Rabiner89-ATO}
%   L.\ R.\ Rabiner,
%   ``A tutorial on hidden Markov models and selected applications in speech recognition,''
%   \textit{Proceedings of the IEEE}, vol.~77, no.~2, pp.~257-286, 1989.
% \bibitem[3]{Hastie09-TEO}
%   T.\ Hastie, R.\ Tibshirani, and J.\ Friedman,
%   \textit{The Elements of Statistical Learning -- Data Mining, Inference, and Prediction}.
%   New York: Springer, 2009.
% \bibitem[4]{YourName17-XXX}
%   F.\ Lastname1, F.\ Lastname2, and F.\ Lastname3,
%   ``Title of your INTERSPEECH 2020 publication,''
%   in \textit{Interspeech 2020 -- 20\textsuperscript{th} Annual Conference of the International Speech Communication Association, September 15-19, Graz, Austria, Proceedings, Proceedings}, 2020, pp.~100--104.
% \end{thebibliography}

\end{document}